# Quantitative Modeling of Point Defects in β-$Ga_2O_3$ Combining Hybrid Functional Energetics with Semiconductor and Processes Thermodynamics

*K.A. Arnab[1], M. Stephens[1], I. Maxfield[1], C. Lee[2], E. Ertekin[2,3], Y.K. Frodason[4], J.B. Varley[5], and M.A. Scarpulla[1,6]*

[1.] Department of Materials Science and Engineering, University of Utah, Salt Lake City, Utah 84112, USA

[2.] Department of Mechanical Science and Engineering, University of Illinois at Urbana-Champaign, Urbana, Illinois 61801, USA

[3.] Materials Research Laboratory, University of Illinois at Urbana-Champaign, Urbana, Illinois 61801, USA

[4.] Centre for Materials Science and Nanotechnology Physics, University of Oslo, Oslo, Norway

[5.] Lawrence Livermore National Laboratory, Livermore, California 93106, USA

[6.] Department of Electrical and Computer Engineering, University of Utah, Salt Lake City, Utah 84112, USA

## Abstract

β-gallium oxide (β-$Ga_2O_3$) is of high interest for power electronics because of its unique combination of melt growth, epitaxial growth, n-type dopability, ultrawide bandgap, and high critical field. Optimization of crystal growth processes to promote beneficial defects and suppress harmful ones requires accurate quantitative modelling of both native and impurity defects. Herein we quantitatively model defect concentrations as a function of bulk crystal growth conditions and demonstrate the *necessity* of including effects such as bandgap temperature dependence, chemical potentials from thermochemistry, and defect vibrational entropy in modelling based on defect formation energies computed by density functional theory (DFT) with hybrid functionals. Without these contributions, grossly-erroneous and misleading predictions arise, e.g. that n-type doping attempts would be fully compensated by Ga vacancies. Including these effects reproduces the experimental facts that melt-grown Sn-doped β-$Ga_2O_3$ crystals are conductive with small compensation while annealing the same crystals in $O_2$ at intermediate temperatures renders them insulating. To accomplish this modeling, we developed a comprehensive modelling framework (KROGER) based on calculated defect formation energies and flexible thermodynamic conditions. These capabilities allow KROGER to capture full and partial defect equilibria amongst native defects and impurities occurring during specific semiconductor growth or fabrication processes. We use KROGER to model 873 charge-states of 259 defects involving 19 elements in conditions representing bulk crystal growth by edge-fed growth (EFG) and annealing in oxygen. Our methodology is transferrable to a wide range of materials beyond β-$Ga_2O_3$. The integration of thermodynamic and first-principles modelling of point defects provides insight into optimization of point defect populations in growth and processing.

Keywords: Gallium oxide, β-$Ga_2O_3$, defects, temperature dependence, growth processes, thermochemistry, vibrational entropy

* Corresponding Author Email: *mike.scarpulla@utah.edu*

## Introduction

β-gallium oxide (β-$Ga_2O_3$) is of intense current interest for power electronics because of its ultra-wide bandgap, high critical field, controllable n-type doping, and the availability of native melt-grown substrates [1–3]. The properties of bulk single crystals and epitaxial layers are intimately tied to point defects and complexes, necessitating predictive models for defect concentrations resulting from varying impurities and processing histories. While density functional theory (DFT) calculations have offered critical qualitative insights into prevalent defects based on calculated formation energies [4–11] translating these findings to actual concentrations of defects present for specific real-world crystal growth and processing requires incorporation of multiple temperature, atmosphere and pressure dependent factors. For example and as we will detail below, a typical calculation holding DFT-calculated formation energies of native and Sn-related defects constant versus temperature predicts that Sn-doped crystals grown at $T_{melt}$ = 2068 °C by edge-fed growth (EFG) with $p_{O2}$=0.02 atm should yield crystals fully-compensated by gallium vacancies ($V_{Ga}$) and related complexes ($Sn_{Ga}$-$V_{Ga}$, $V_{Ga}$-$V_O$, etc.). Yet Sn-doped wafers can be purchased commercially and in actuality display (nearly) degenerate n-type doping with <1% compensation [12,13]. The accuracy of DFT-based defect modelling of charge transition levels compared to experiments brought by large supercells with finite-size correction schemes, hybrid functionals and self-interaction corrections usher in a new era in defect modelling in which it is worth the effort to incorporate the aforementioned thermodynamic effects [14,15]. The case study of systematically incorporating a suite of thermodynamic contributions for β-$Ga_2O_3$ herein sets a benchmark for achieving quantitative agreement with experiments by combining all of these factors [16,17].

Here we utilize a comprehensive, transferrable, quantitative framework dubbed KROGER [18] for modelling full, partial, and constrained defect equilibria to emulate defects resulting from real-world growth and processing [19]. KROGER [18], named after F.A. Kroger who gave exhaustive treatments of point defect concentrations [20,21], allows us to take a set of DFT-computed formation energies for the charge states of defects and complexes and compute their numbers given specified thermodynamic conditions representing real-world processing [14]. β-$Ga_2O_3$ is used as an exemplar and through this modelling we elucidate new insights specific to this material's defect chemistry, but of course the methodology is transferrable to other materials. We account for the temperature dependencies of the bandgap and densities of states, temperature- and pressure-dependent chemical potentials from the Ga-O binary system, degenerate statistics, self-trapped holes (STHs), and a minimal quantum harmonic oscillator model for vibrational entropy change for defects which can exceed 1.5 eV/defect near $T_{melt}$. We model EFG-grown unintentionally-doped (UID), Sn-, Fe- and Mg-doped crystals by calculating defect equilibrium based on their reported total dopant and impurity concentrations rather than assuming fixed chemical potentials. We illustrate how assuming kinetic trapping of dopants in combination with the aforementioned temperature dependencies yields agreement with experiments. The ability to model various constrained equilibria gives KROGER added capabilities for modelling semiconductor processing beyond the state of the art [22–29].

We illustrate the scalability of our modelling framework and ability to explicitly handle trace impurities by including a total of 19 elements distributed across 259 defects having 873 charge states, as well as self-trapped holes localized on $O_I$ and $O_{II}$ atoms[30,31]; limited only by the availability of a self-consistent set of DFT-calculated formation energies for all these defects' charge states. Such calculations at dozens of temperatures run within minutes on a personal computer, allowing rapid exploration of various effects. We accurately model defects in Sn, Fe, and Mg-doped edge-fed growth (EFG) crystals, Bridgeman crystals grown under different $p_{O2}$, and annealing experiments in pure $O_2$ at 1 atm. We compare our results with the well-established parameters of Sn-doped $Ga_2O_3$, such as the electron and donor concentration ratio and DLOS data on $V_{Ga}$ -related defect densities [32]. This work provides novel insights into defects and complexes in β-$Ga_2O_3$ while also illustrating the capabilities of KROGER which was built for generality and transferability to defect systems in other materials. This comprehensive approach yields insights beyond those typically possible from DFT defect formation energetics evaluated in certain rich and poor chemical potential conditions.

## Results & Discussion

*Establishing a Baseline Model for Defects in Donor-Doped EFG β-$Ga_2O_3$*

In this paper, we primarily focus on the EFG method [33], as the properties of commercial wafers grown by this method are consistent and well-characterized. We also provide some results in the Supplementary Materials for Bridgman[34] processes which utilize different noble metal crucibles. Both EFG and Bridgman growth of β-$Ga_2O_3$ utilize 1 atm total pressure, with EFG using 2% oxygen $pO_2$ and Bridgman 20% $pO_2$ [35,36]. All of the melt-growth techniques occur at $T_m$=2068 K, and despite the different $pO_2$ values converge on very close predictions because for gasses, chemical potential depends logarithmically on pressure. Additionally, we have carried out annealing experiments on 500-700 μm thick Sn-doped wafers in 1 atm $pO_2$ at 1300-1400 K for durations up to 2 weeks, which was the time required for elimination of free carrier absorption.

Figure 1 illustrates three different models of temperature-dependent defect concentrations for EFG-grown Sn-doped β-$Ga_2O_3$ resulting from different assumptions about chemical potentials and impurity concentrations. To simplify plots, we have grouped similar defects and complexes, as discussed in the Supplementary Materials. Figure 1(a) corresponds to kinetic trapping of [Sn]=4.5x10$^{18}$ /cm$^3$, [Si]=2x10$^{17}$ /cm$^3$ and [Fe]=7x10$^{16}$ /cm$^3$ and O and Ga chemical potentials fixed by Eqs. 2 & 7 (see in Computational Methods) with $p_{O2}$=0.02 atm. Subfigure (b) presents results assuming the O and Ga chemical potentials do not vary with temperature for O-rich conditions, in other words for the "O-rich" conditions presented in Varley et al.[8–10] and Frodason et al.[4–6]. Part (c) is the same as (a) but with $\Delta\mu_{Sn}$ set by equilibrium with $SnO_2$ using measured thermochemical data. The different temperature dependencies of $\Delta\mu_{Ga}$, $\Delta\mu_O$, and $\Delta\mu_{Sn}$ for these cases are plotted in Fig. 2. In Fig. 1 (a) and (c), the total concentrations of other impurities are fixed at representative values from Kuramata et al.[33] In (a) and (c) the temperature dependence of bandgap is taken from Lee et al.[37] with a representative assumption of 40% of $\Delta E_g(T)$ occurring in the conduction band, as discussed in depth below in the context of Fig. 4. While not shown here, it is rather a simple task to combine models like (a) and (c) to make models accounting for dopant equilibration with 2$^{nd}$ phases at high temperatures but freezing in at some intermediate temperature.

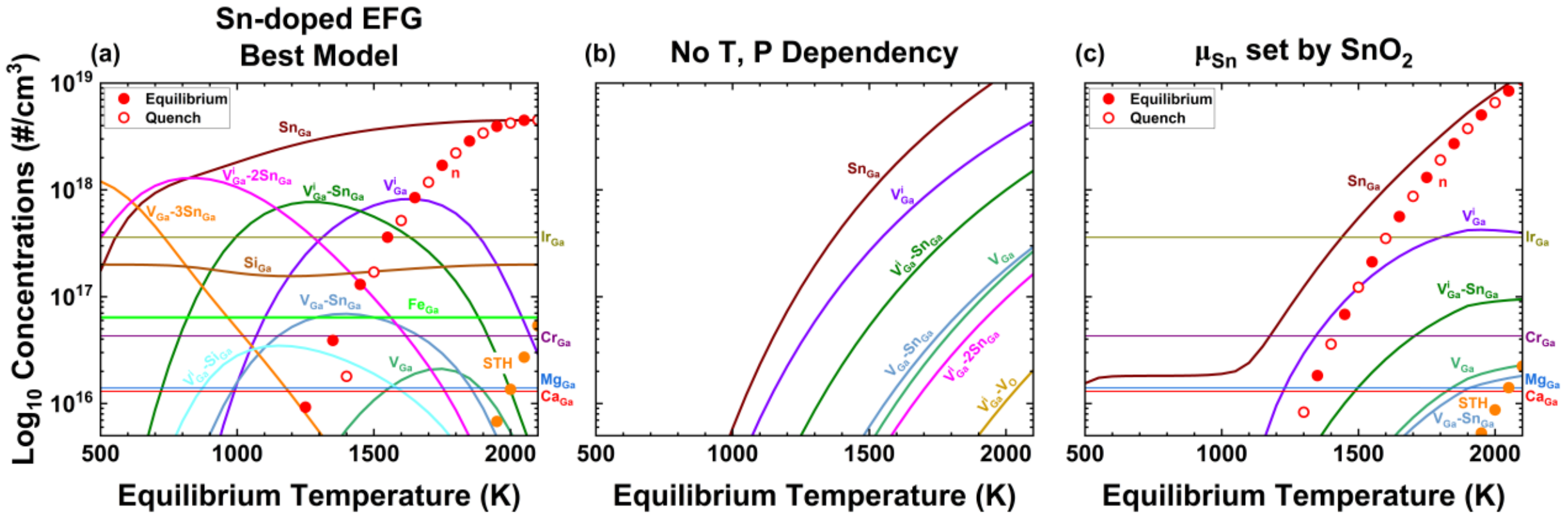

**Figure 1.** Defect concentrations calculated for Sn-doped β-$Ga_2O_3$ in different scenarios. (a) Fixed concentrations of Sn and other impurities and incorporating the full set of temperature-dependent effects for $p_{O2}$=0.02 atm used in EFG growth[33]. Both equilibrium and quenching scenarios are consistent with Sn-doped wafers being n-type with <1% compensation as long as the native defect

system freezes-in by ~1950 K. Overlap of equilibrium and quenched results predicts insensitivity to cooling rate. (b) "O-rich, Sn-doped" conditions[38] without any temperature dependencies *predicting that Sn-doped wafers should be insulating with >99% compensation (n in the $10^{15}$ range) for all temperatures, which disagrees with reality*. (c) Similar to (a) in terms of other impurities but with $\mu_{Sn}(T)$ set by equilibrium with $SnO_2$. In this scenario, agreement with real EFG Sn-doped crystals c*ould* occur only if the native defect system and [Sn] simultaneously freeze-in at ~1850 K, which would be more likely for larger crystals. This would be expected to depend sensitively on cooling rate, which is improbable given the widely-reported ease of reliable, low compensation bulk doping with Sn. Note that (c) represents the temperature-dependent solubility limit for Sn in β-$Ga_2O_3$ from thermochemistry; the fact that [Sn]>$10^{18}$ /cm$^3$ is achievable without $SnO_2$ precipitation is direct evidence for kinetic trapping of Sn. Note, [Si] was omitted in (c) to allow Sn to be the dominant donor for all temperatures and that the equilibrium solubility becomes dependent on the residual [$Mg_{Ga}$]+[$Ca_{Ga}$]-[$Zr_{Ga}$] at low temperatures.

We conclude that the model in (a) is the most likely to represent real-world crystals because it agrees with the physical expectation that the diffusion of Sn will freeze in somewhere near $T_m$ when we consider large crystals having appreciable dimensions (multiple centimeters) and that three important experimental observations are satisfied. These are 1) that real-world EFG-grown Sn-doped crystals are nearly degenerately doped with negligible compensation such that [Sn]=n within experimental limits of a few %; 2) that annealing in $p_{O2}$=1 atm at 1300-1400 K will change such samples towards insulating as shown in Figure 3 is widely observed [39]; and 3) that the concentration of compensating $V_{Ga}$-related defects and complexes responsible for the $E_c$-2 eV signal measured using deep level optical spectroscopy (DLOS) is in the low $10^{16}$ /cm$^3$ range. We note that points 3) and 1) are related since $V_{Ga}$ and their complexes are the dominant compensating native defects. The suppression of $V_{Ga}$ and related complexes at intermediate to high temperature is driven primarily by the strong temperature dependence of $E_g$, as shown in Fig. 4 and the surrounding discussion. As shown further herein and in the Supplemental Material, a limited range of models can yield such agreement with the 3 constraints mentioned. Model (b), which represents the O-rich point in Fig. 2(b), clearly disagrees with as-grown real-world Sn-doped wafers since it predicts that >99% of the $Sn_{Ga}$ donors in Sn-doped wafers grown by EFG should be compensated with $V_{Ga}$ and thus Sn-doped wafers should be insulating regardless of the temperature at which defects freeze in. Model (c) represents the extension of typical phase boundary mapping of allowed chemical potential ranges but using measured thermochemical data and for experiments carried out at constant $p_{O2}$ vs temperature (as opposed to constant $\Delta\mu_O$). Note that the electron concentration follows [Sn] closely down to the temperature where [Sn]≈[Mg] then it starts to decrease again following $V^i_{Ga}$-$Sn_{Ga}$, since Ir offers a deep donor transition and Cr offers no transitions in the upper part of the bandgap. We consider it possible but improbable that the scenario of part (c) represents actual samples. This is because 1) the [Sn] in the melt is intentionally kept below the solid solubility limit and 2) no evidence of precipitation of $SnO_2$ or macroscopic fluctuations in [Sn] and n depending on position or cooling rate have been reported for bulk crystals of β-$Ga_2O_3$:Sn.

In Fig. 1(a), the constraint that total [Sn] is constant manifests as the dominant Sn-containing defect or complex changing as temperature is lowered. At high temperatures, isolated $Sn_{Ga}$ are the dominant species, then below ~1300 K complexes with $V_{Ga}$ begin to condense in order of decreasing maximum negative charge state with $V_{Ga}$-$3Sn_{Ga}$ having only a neutral charge state for n-type doping and related complexes being stable at room temperature. We note that complexes containing 2 and 3 Sn would require significant Sn diffusion over distances of ~ 6 nm (the mean Sn-Sn distance at mid-$10^{18}$ /$cm^3$ concentration) at these lower temperatures which would be kinetically hindered during boule cooling[5]. In this work we neglect these effects, but we plan to incorporate future improvements into KROGER to account for such defect kinetics. Similar trends for the Si known to provide unintentional doping in most sample types is also seen, but for Fe and other impurities known to be in EFG-grown crystals the stability of single charge states dominate for all temperatures such that they appear as horizontal lines. At the very highest temperatures [39] near $T_m$, $V_O$ and related complexes offering ~1 eV ionization energies provide equilibrium n>[Sn] (but at room temperature are not ionized). This encourages formation of compensating acceptor defects that, when frozen in, result in lower $N_D$-$N_A$ at room temperature. To state this plainly, although their donor levels are too deep to be ionized at room temperature, the behavior of $V_O$ at high temperature influences [$V_{Ga}$], and thus indirectly does influence n at room temperature depending on where [$V_{Ga}$] freezes in which in turn depends upon cooling rate, sample dimensions via diffusion, and the kinetics of complexes. We also investigated the effects of unintentional hydrogen up to densities ~$10^{17}$ /$cm^3$ (the upper bound estimated to be unintentionally incorporated into CZ-grown crystals[39,40] ). We found that this level of hydrogen only minorly changes the results for Sn-doped as-EFG-grown crystals and for $O_2$ annealing (in which H would all be removed)[41]; these results are shown in the Supplemental Material. Finally, we tested the effects of STHs – near the melting point accounting for STHs results in much higher intrinsic carrier density such that carriers rather than defects may dominate charge balance, which again can modify the prediction of dominant defects frozen in during processing. The including or excluding STHs did not change any qualitative results, although they do tend to promote the formation of acceptors like $V_{Ga}$ and slightly modify the details of predicted carrier density for samples annealed in $O_2$ at 1000-1100°C.

We do note that our method of estimating $\Delta S_{vib}$ significantly penalizes vacancy complexes such as $V_O$-$V_{Ga}$ divacancies (see Figure 5). Since changes in mode frequencies may mitigate this effect, we consider our predictions for these complexes at the highest temperatures to carry some increased uncertainty. The computational costs for DFT calculations using hybrid functionals of defective supercells large enough to approach the dilute limit are nearly prohibitive; when high quality calculations become available KROGER can incorporate this effect. Fortuitously, this effect will be largest at high temperatures above which impurities and defects equilibrate for slow growth rates and will decrease in importance linearly with temperature. For these reasons we expect that our predictions for temperatures below which the defect system is frozen-in are on rather sound footing.

Summarizing, we find that agreement of defect modelling for β-$Ga_2O_3$ with real-world observations requires including at least 1) kinetic trapping of dopants like Sn rather than assuming equilibrium with competing phases, 2) temperature dependent chemical potentials taken from thermochemistry, and 3) temperature dependent band edges. The other effects discussed help to tune the details of the results herein, but may be more important for other materials and growth situations.

*Importance of "Real-World" Chemical Potentials*

We now proceed to analyze and discuss the contributions of these temperature-dependent factors and related assumptions about the thermodynamic environment on predicted defect concentrations. Figure 2(a) shows the differences in chemical potentials vs temperature for oxygen and gallium under different thermodynamic assumptions. The black solid (dashed) lines show $\Delta\mu_O$ ($\Delta\mu_{Ga}$) for β-$Ga_2O_3$ in equilibrium with "O-rich" conditions discussed in the papers reporting our set of defect formation energies[4–6,8–10]; this corresponds to holding the chemical potentials and formation energy of $Ga_2O_3$ ($G°_{Ga2O3}$) constant vs. temperature. Of course, in experiments it is much more common to hold the oxygen partial pressure constant since $O_2$ is a gas. This results in a roughly linearly-decreasing dependence on T which can be estimated as -1 eV per 700 K. The solid red (solid blue) curves show the experimental temperature dependencies for holding $p_{O2}$=0.02 atm constant as in EFG growth and $p_{O2}$=1 atm as during annealing experiments in a tube furnace[42,43]. The $\Delta\mu_{Ga}$ required to maintain $Ga_2O_3$ in equilibrium (neither growing nor decomposing) is shown as dashed red or blue lines, taking into account also the temperature dependence of $G°_{Ga2O3}$, which is driven by its specific heat capacity ($C_p$).

The dash-dot green series of Fig. 2(a) shows the temperature dependence of the maximum $\Delta\mu_{Sn}$ allowable without precipitation of $SnO_2$ for $p_{O2}$=0.02 atm calculated analogously to $\Delta\mu_{Ga}$. Assuming the concentration of Sn is held constant at [Sn]=4.5x10$^{18}$ /cm$^3$ for $p_{O2}$ = 0.02 atm as for Fig. 1(a) produces the non-monotonic solid green curve. This dependence arises from self-consistently solving for both $\Delta\mu_{Sn}$ and $E_F$ at each temperature thus is a function of both phase thermochemistry and the set of defects and complexes included in the calculation. The crossing point of the green solid and dash-dot curves near 1850 K represents the temperature at which the Sn solubility limit falls below 4.5x10$^{18}$ /cm$^3$, as can be seen in Fig. 1(c). In EFG crystal growth of Sn-doped substrates, the concentration of Sn added to the melt is less than the solubility limit in the solid at $T_m$, thus at high temperatures [Sn] in the solid is limited by supply in the melt but as the crystal cools the Sn is kinetically trapped at concentrations exceeding the solubility limit simply because the diffusion kinetics become too slow to precipitate 2$^{nd}$ phases.

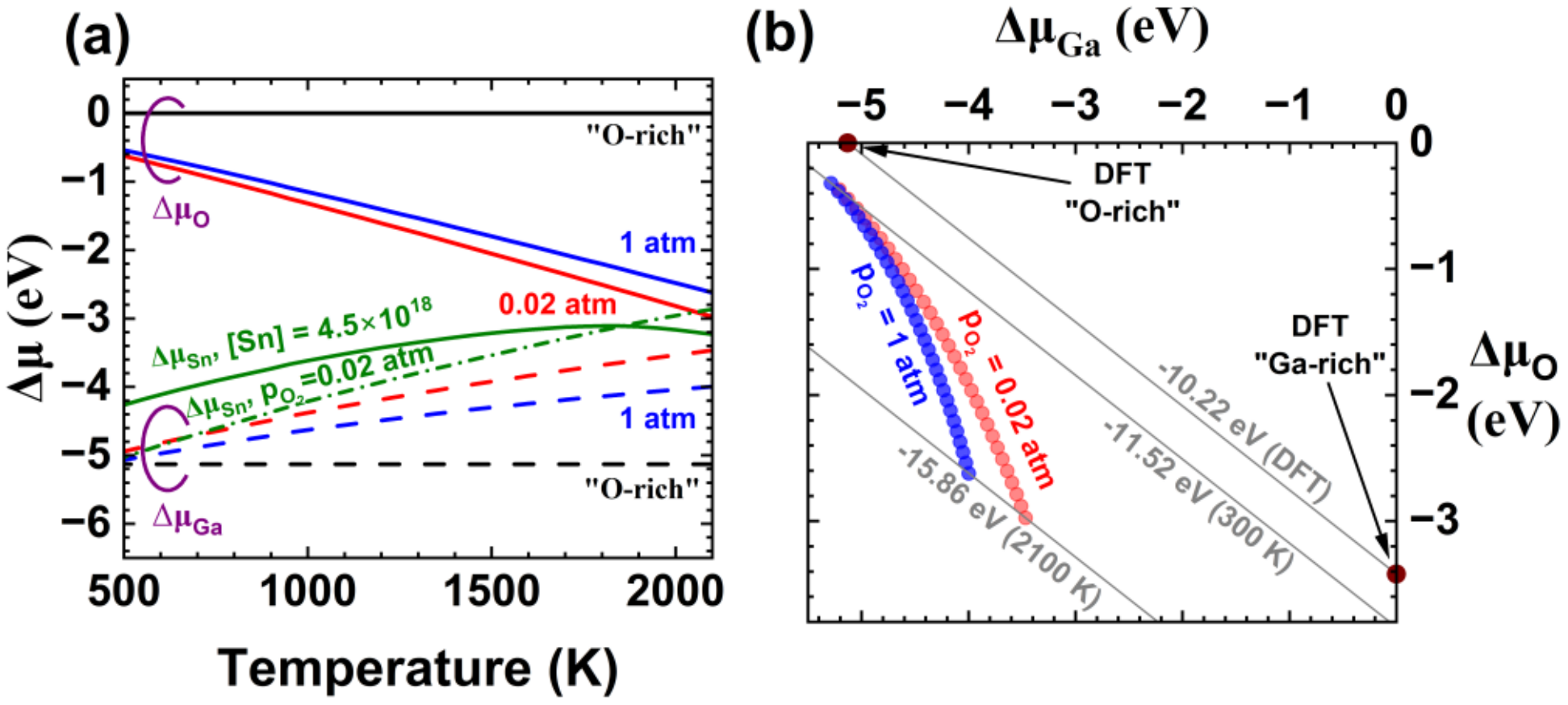

**Figure 2.** (a) Temperature dependent chemical potentials for Ga, O and Sn under different thermodynamic scenarios with $\Delta\mu_O$ solid and $\Delta\mu_{Ga}$ dashed: typical O- or Ga-rich (black), based on thermochemical functions from the Ga-O system for $p_{O2}$=0.02 and 1 atm (red, blue). For Sn

(green), solid denotes the case where the total [Sn] is fixed while dash-dot is derived from the $SnO_2$ phase boundary for $p_{O2}$=0.02 atm. (b) Plot of $\Delta\mu_{Ga}$ and $\Delta\mu_{O}$ in Ga-O chemical potential space. At 0 or 300 K, the experimental $\Delta H_F = \Delta G^o$ (grey diagonal lines) is within 15% of the DFT-calculated value, but as $T_m$ is approached this error increases to 50% and the locus of conditions for which $p_{O2}$ is held constant moves quite far from either O-rich or Ga-rich conditions (blue or red circles). This illustrates why using chemical potentials from thermochemistry gives a more realistic prediction for defects present.

Figure 2(b) translates Fig. 2(a) into chemical potential space by parametrically plotting $\Delta\mu_{Ga}$ and $\Delta\mu_{O}$ at each temperature. O-rich and Ga-rich conditions are single points, shown for the DFT-calculated $\Delta H_F$ = -10.22 eV/FU[5]. The diagonal line connecting these two points gives all combinations of $\Delta\mu_{Ga}$ and $\Delta\mu_{O}$ satisfying $2\Delta\mu_{Ga}+3\Delta\mu_{O}=\Delta H_F$ and moving between the endpoints would correspond to traversing the single-phase β-$Ga_2O_3$ field on the T-x phase diagram while remaining in equilibrium. Analogous lines based instead on experimental thermochemistry are presented for 300 and 2100 K; it can be seen that the standard DFT result is a minor underestimate at 300 K but that the discrepancy approaches 50% at high temperatures because of the neglect of phonon specific heat capacity. We note that crystal growth requires at least a slight departure up and right from the equilibrium lines in order to provide a driving force; analogously, decomposition/etching occurs for conditions slightly down and left. The blue (red) data series parametrically plot the temperature dependence of Ga and O chemical potentials for β-$Ga_2O_3$ in equilibrium with $p_{O2}$=1 atm (0.02 atm), in correspondence with Fig. 2(a). These do not fall on one diagonal line since $G^\circ_{Ga2O3}$ becomes increasingly negative (by >4 eV/FU) from 300-2100 K. While the locus ($\Delta\mu_{Ga}$, $\Delta\mu_{O}$) for 1 atm pure $O_2$ is <15% away from "O-rich conditions" at 0 or 300 K, at elevated temperatures constant $p_{O2}$ experiments have loci many eV away from either O-rich or Ga-rich conditions, translating to enormous differences in predicted defect concentrations (factors >$10^7$ !). Achieving the "O-rich" conditions presented in DFT papers during bulk crystal growth from the melt would require 1) enormously elevated $p_{O2}$, 2) use of alternate O sources such as ozone[44,45], or 3) dissociation and or excitation of $O_2$ molecules as in plasma-assisted molecular beam epitaxy (PAMBE). Summarizing, real-world bulk crystal growth or processing conditions are probably quite far from "O-rich" or "Ga-rich" conditions presented in many DFT computation papers, thus large differences in concentrations of defects predicted are to be expected when one or more components of the crystal is not a condensed phase. Growth methods using excited reactants can have higher chemical potentials, but their chemical potential values are difficult to quantify.

Figure 3 investigates the effect of assumptions about O chemical potential while holding [Sn]=$4.5x10^{18}$ /cm$^3$ for (a) O-rich conditions, (b) $p_{O2}$=0.02 atm, $p_{tot}$=1 atm and (b) $p_{O2}$=$p_{tot}$=1 atm. To clarify the effects, all native defects were included in calculations but Sn is the only impurity included. The solid red circles show the free electron density n at the given temperature, while open circles represent predicted results for instantaneous quenching from each temperature to 300 K. It is immediately clear from Fig. 3(a) that the assumption of O-rich conditions at crystal growth and cooling temperatures would yield a prediction that Sn-doped crystals should be insulating rather than the commonly-experienced nearly degenerate doping with nearly unmeasurable compensation. Using experimental thermochemical data for $p_{O2}$=0.02 or 1 atm as shown in (b) and (c) results in predictions for both equilibrium and quenching compatible with that real-world observation as long as the defect system freezes-in (for large crystals) somewhere between $T_m$ and ~1950 K. Please note that the transition temperature at which the ratio n/[$Sn_{Ga}$] begins to deviate below 1 changes with the assumption of how much of the temperature dependent bandgap occurs in the conduction band, as discussed in detail in the following section. Figure 3 assumes 40% (f=0.40), but plots for f between 0 and 1 are shown in the Supplemental Material. As noted before, we suspect that the native defect system probably freezes-in at the lower end of this range given that DLOS finds $V_{Ga}$-related defects in the low $10^{16}$ /cm$^3$ range[32] and annealing above ~900 °C is required to observe changes in conductivity of doped samples and diffusion of substitutional impurities[42,43,46–48]. We have performed (unpublished) experiments annealing Sn-doped wafers in 1 atm $O_2$ at 1000-1100 °C. Samples become transparent to the eye and the free carrier absorption becomes undetectable using Fourier transform infrared (FTIR) after 2 weeks. Comparing Fig. 3 (b) and (c), this effect is predicted fairly well since the predicted n(T) for both equilibrium and quenching cases is lower at all temperatures for $p_{O2}$=1 atm compared to the 0.02 atm used for EFG growth and reaches the $10^{16}$ /cm$^3$ range 1300-1400 K. However, our FTIR experiments indicate that the true carrier density is probably lower, thus some further improvements in our modeling such as details of $\Delta S_{vib}$, fraction f of $E_g(T)$ in the conduction band, hydrogen incorporation, and freeze-out of different defects at different temperatures should be investigated more fully in the future.

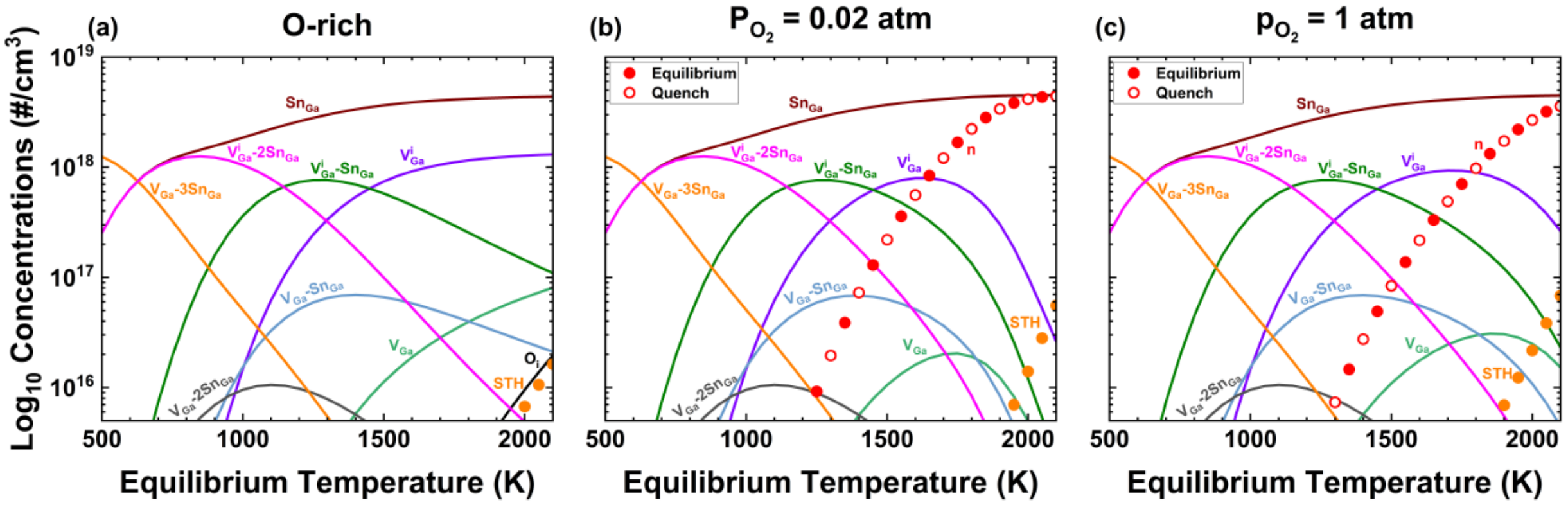

**Figure 3.** Calculated concentrations of defects and holding [Sn] = 4.5 $\times 10^{18}$ constant for (a) O-rich conditions holding $\Delta\mu_O$ and $\Delta\mu_{Ga}$ as well as $E_g$ constant vs. T as is typical in much existing literature. Note that the predicted carrier density is not even visible, thus samples would be predicted to be insulating. (b) for EFG crystal growth ($p_{tot}$ = 1 atm, $p_{O2}$ = 0.02 atm, f = 0.40), and (c) $O_2$ annealing ($p_{tot}$=$p_{O2}$=1 atm, f = 0.40). Solid circles represent equilibrium concentrations, while open circles show results for quenching from each equilibration temperature to 300 K.

*Importance of Temperature-Dependent Bandstructure*

Having examined the effects of different thermodynamic and kinetic constraints on the elements incorporated into the modeling, we now turn attention to the effects of temperature dependence of the bandgap, which in β-$Ga_2O_3$ decreases by nearly 2 eV from 300 K to $T_m$ without even accounting for STHs [37]. Typical experiments like optical absorption cannot resolve the absolute band edge positions independently; only their difference $E_g(T)$ is accessible. Analogously, DFT using pseudopotentials that exclude deep core levels can result in ambiguity in that different correction schemes for electric potential self-interactions within the supercell may result in conflicting predictions[37]. We thus have investigated a range of scenarios in which the temperature dependent change in bandgap ($\Delta E_g$) is apportioned as

$$\Delta E_g(T) \equiv E_g(T) - E_g(0) = -f[E_C(T) - E_C(0)] - (1-f)[E_V(T) - E_V(0)] \qquad (1)$$

in which $E_C$ ($E_V$) is the absolute conduction (valence) band energy and f represents the fraction of the total change in bandgap occurring in the conduction band. We note that the lighter-mass oxygen atoms comprising the valence band edge states do appear to dominate the overall band gap reduction at higher temperatures, and that smaller *f* values (e.g. f=0.40) are focused on more below.

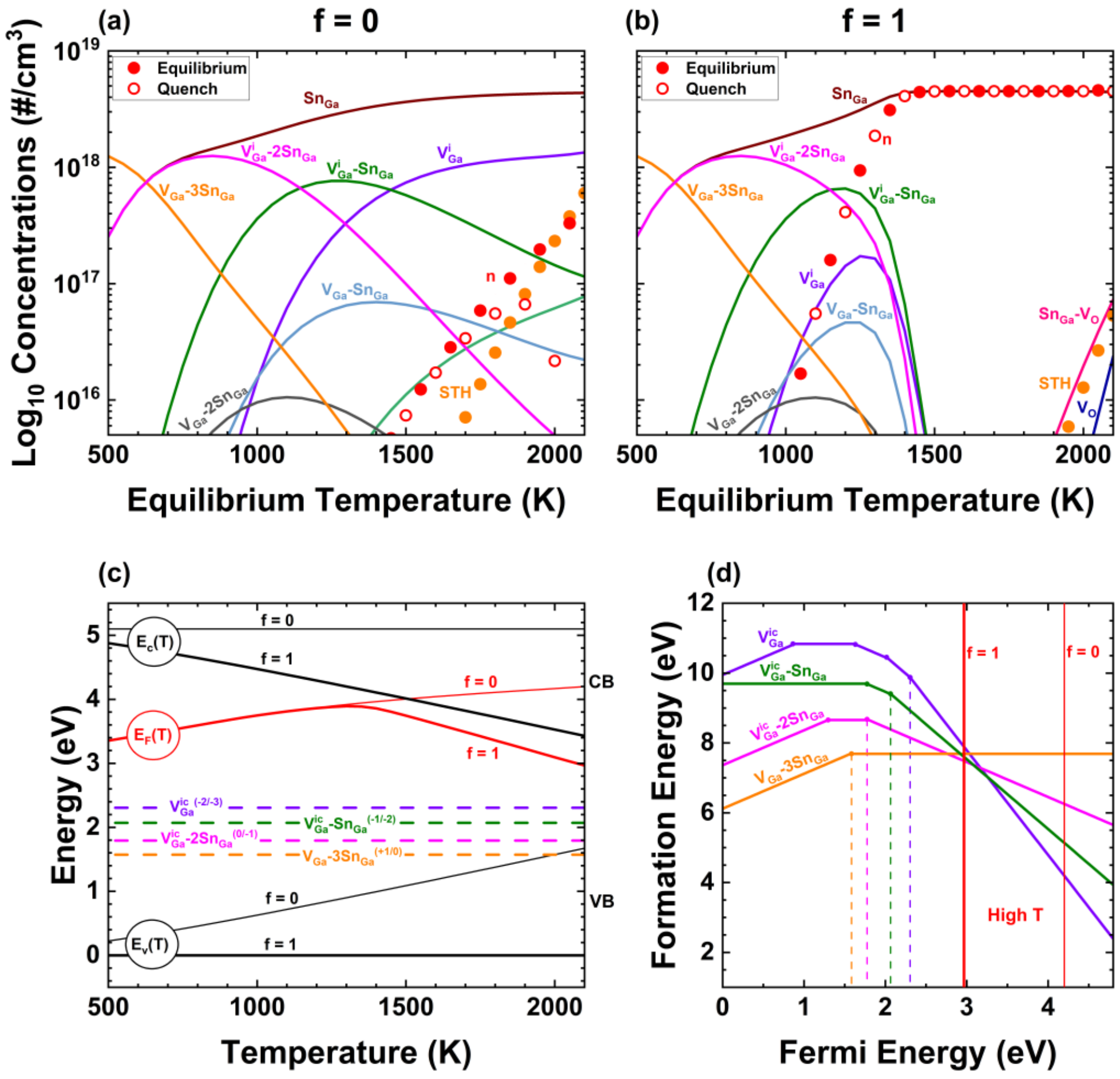

**Figure 4.** Effects of band edge variation with temperature for conditions modeling EFG of Sn-doped wafers ($p_{O2}$=0.02 atm, [Sn]=4.5x$10^{18}$ /cm$^3$) assuming (a) f=0 such that only the valence band edge varies or (b) f=1 such that only the conduction band edge varies. If the conduction band and Fermi energy move towards mid gap, the dominant $V_{Ga}$ native acceptor-like defects are dramatically suppressed. (c) Plots of band edges for f=0 or 1 along with the highest charge transition levels for $V_{Ga}$ and its complexes with Sn, and Fermi levels versus temperature. For f=1 and intermediate temperatures, $E_F$ is pushed downwards at nearly the same rate as $E_C$ (~1 eV/1000 K) resulting in 3, 2, and 1x times increases in formation energies for the most negative charge states of $V_{Ga}$, $V_{Ga}$-$Sn_{Ga}$, and $V_{Ga}$-$2Sn_{Ga}$ complexes respectively, as seen in (d). (d) depicts the most stable charge states of these defects and complexes versus $E_F$ along with the $E_F$ values near $T_m$ from part (c).

Figure 4 shows the dramatic effect exerted by the conduction band's temperature dependence on $V_{Ga}$ and its complexes, which are the dominant native defects that compensate shallow n-type dopants. In (a) and (b), we show computed defect concentrations representing EFG growth of crystals doped with [Sn]=4.5x10$^{18}$ /cm$^3$, assuming $\Delta E_g(T)$ occurs entirely in the valence band (f=0) in (a) and entirely in the conduction band (f=1) in (b). The conditions are the same as in Fig. 1(a). Since $E_F$ is in the top half of the bandgap for n-type doping, changes in $E_V$ with temperature have no effect on the net doping $N_D$-$N_A$ and charge balance demands large concentrations of compensating $V_{Ga}$ and complexes with Sn. However, changes in $E_C$ with temperature dramatically suppress these native defects. This is one of the strongest effects we infer must be occurring in real crystals; without it, Sn-doped wafers grown by EFG should be heavily compensated as in (a) and DLOS experiments should reveal $E_C$-2 eV trap numbers approaching 10$^{18}$ /cm$^3$. Large values of f suppress $V_{Ga}$ compensating defects sufficiently for the appearance of a wide temperature range over which the defect system can freeze-in and result in n=[Sn] at room temperature (Supplemental Material). Larger values of f (>0.50) predict less sensitivity to the freezing-in temperature of compensating defects for Sn-doped EFG crystals and thus the as-grown behavior. However, smaller values of f are required to simultaneously give agreement that 1300-1400 $^oC$ $O_2$ annealed samples should be insulating – thus the annealing experiments provide more stringent constraints on the overall model. The Supplemental Material shows cases for $p_{O2}$=1 atm for f values of 0, 0.5 and 1. We determined our most-likely estimate for f=0.40 (+/- 0.05 or so) as the value that best satisfies the two experimental constraints simultaneously (given all the other model assumptions). Lee et al. [37] concluded from DFT calculations that most of the temperature dependence of the bandgap occurs in the valence band, although an alternate absolute potential alignment scheme discussed in that work predicted the opposite. Figure 4(a) shows that setting f=0 is inconsistent with conducting EFG Sn-doped samples, which is one of the main experimental observations. Thus, our modelling provides circumstantial evidence for the valence band contributing slightly more than half, but direct experimental investigations using e.g. X-ray techniques would be desirable. Our mode-counting of simple quantum oscillators treatment of $\Delta S_{vib}$ suppresses complexes including $V_O$ at high temperature, which will affect the predicted compensation if crystals were quenched rapidly. Figure 5 delves further into the range of models yielding agreement with experimental observations.

Figure 4 (c) shows the temperature dependent band diagram including Fermi energy corresponding to parts (a) and (b) and, with (d), explains the mechanism behind $V_{Ga}$ suppression. The black thin (thick) lines show $E_C$ and $E_V$ for f=0 (f=1), while red lines show the solution for $\varepsilon_F$ in both cases. For f=1, the fact that we impose [Sn]=constant becomes important, since $E_C - \varepsilon_F = k_B T \ln\left(n/N_C(T)\right)$, in the absence of strong compensation $\varepsilon_F$ will follow the temperature dependence of $E_C$. Examining part (d) makes the mechanism apparent; as $E_C$ and $\varepsilon_F$ shift down at high temperatures (by nearly 2 eV from 0 K for f=1), the formation energies of the dominant $V_{Ga,ic}^{3-}$, $(V_{Ga,ic}\text{-}Sn_{Ga})^{2-}$, and $(V_{Ga,ic}\text{-}2Sn_{GaII})^{1-}$ increase by approximately 6, 4, and 2 eV respectively, resulting in their near-total suppression and prevention of native defect compensation at high temperatures. The case of growth of β-$Ga_2O_3$ from the melt provides an extreme case of the importance of including temperature dependent band edges in defect modelling because of its high $T_m$ and strong electron-phonon coupling[3,37], but we recommend that this effect should be more routinely included in defect concentration modelling. If $\Delta S_{vib}$ and $E^o$ (see Eq. 2 in Computational Methods) differ for different charge states, the charge transition level between them will become temperature-dependent [49]. KROGER has the capability to incorporate such temperature-dependent formation energies, which represent one of the next frontiers in defect energy calculations.

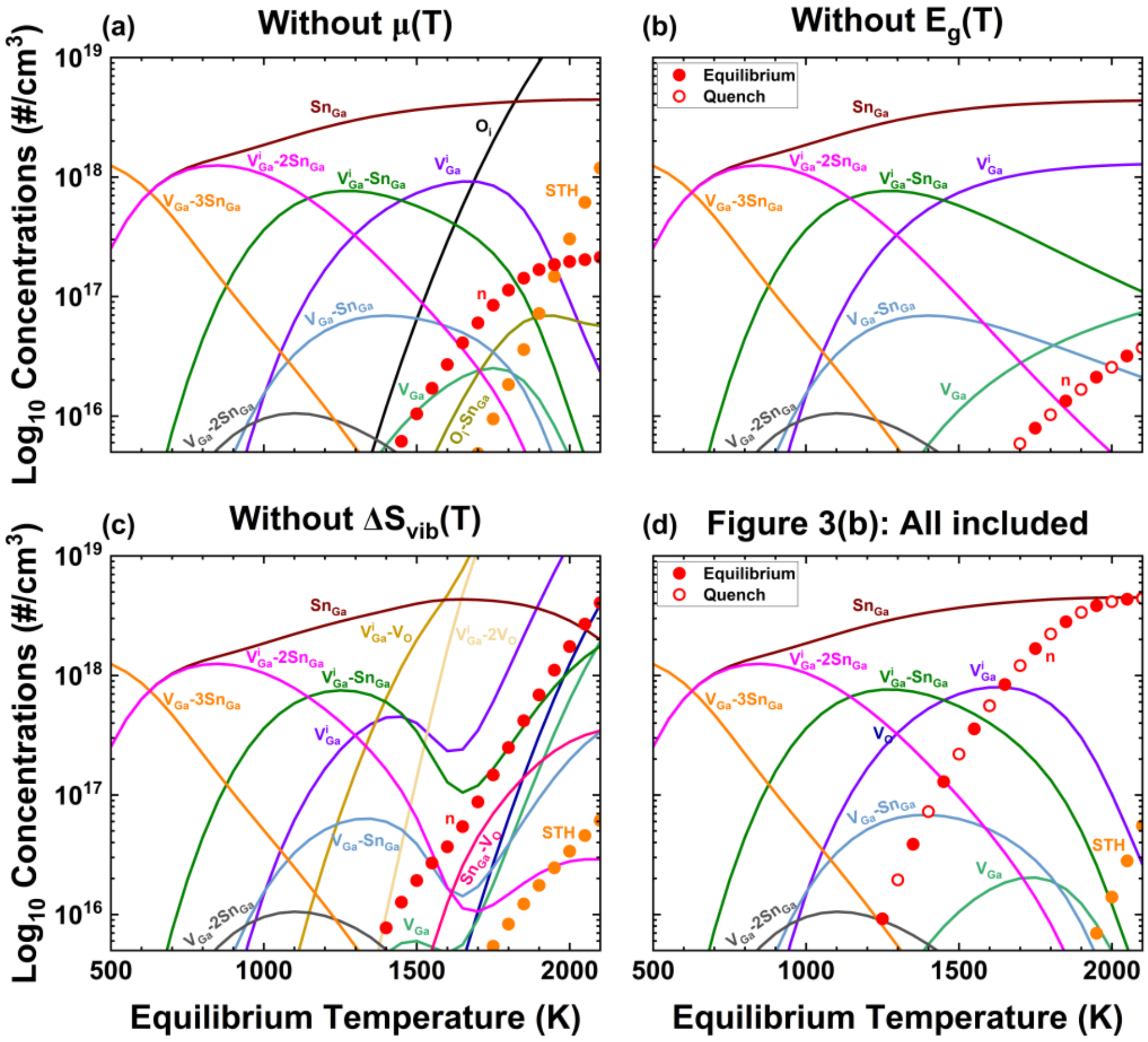

**Figure 5.** Effects of omitting one temperature dependence at a time to investigate their importances. Conditions kept constant are $[Sn]=4.5x10^{18}$ $/cm^3$ and f=0.40. (a) Adopting T-independent O-rich conditions as in Fig. 1(b) predicts $O_i$ are the dominant compensating acceptors at high temperature and at no temperature does n=[Sn]. (b) Effect of $E_g(T)$, $N_C(T)$, and $N_V(T)$ held constant at 300 K values, again there is no temperature where n=[Sn]. (c) Effect of omitting $\Delta S_{vib}(T)$ – here there is only one temperature at which n=[Sn] so coincidence would be required for agreement with experiments. Also, the concentrations of $V_{Ga}$ and divacancy complexes increase to at% concentrations, which would be possible to measure. (d) Replication of Fig. 3(b) including all effects to facilitate side-by-side comparison.

Figure 5 investigates the criticality of including various temperature-dependent parameters – $\Delta\mu(T)$, $E_g(T)$, and $\Delta S_{vib}(T)$ – in achieving agreement with experimental observations. In each subfigure unless noted, $[Sn]=4.5x10^{18}$ $/cm^3$, $p_{O2}$=0.02 atm as for EFG, and f=0.40 are used. Figure 3(b) provides the results when all three temperature-dependent effects are included, and is replicated as Fig. 5(d) to facilitate side-by-side comparison. Figure 5(a) includes $E_g(T)$ with f=0.40 (and the $T^{3/2}$ dependencies of $N_c$ and $N_v$) and the quantum estimate for $\Delta S_{vib}$ but keeps chemical potentials constant at the DFT-estimated 0 K values ($\Delta\mu_{Ga}$=-5.11 eV and $\Delta\mu_O$=0 for O-rich conditions). Despite the fact that the $V_{Ga}$ are suppressed at high temperatures, agreement with reality could not be found as n<<[Si] for all temperatures. Notably also, the unrealistically-high $\Delta\mu_O$ compared even to pure oxygen conditions ((see Fig. 2) causes $O_i$ to provide compensation at high temperatures rather than $V_{Ga}$ related defects including $V_{Ga}$-$V_O$ divacancies[4]. $O_i$ is believed to have extremely low migration energies[50], thus is expected remain equilibrated even at room temperature, however if other compensators like $V_{Ga}$ are suppressed at high temperature because of the presence of $O_i$ but frozen in, the presence of $O_i$ can still change the defects present at 300 K. Thus, it would not be proper to build reduced models in which $O_i$ was omitted. This is one example why we favor avoiding a-priori assumptions regarding dominant defects. KROGER is sufficiently computationally efficient that dominant defects and reduced models may be identified after the fact without significant delays.

Figure 5(b) includes temperature dependent chemical potentials for O-rich conditions (p=0.02 atm pure $O_2$ as in Fig. 2), but assumes a temperature-independent bandgap and density of states values. There is no possibility of agreement with real-world Sn-doped wafers since n<<[Sn] for all temperatures for equilibrium or full quenching. Finally in Fig. 5(c), temperature dependent band parameters and realistic chemical potentials are incorporated but $\Delta S_{vib}$ is omitted. Here agreement with experiment is possible, but only if the defect system happens to coincidentally freeze in at one specific temperature. Notably, concentrations of $V_{Ga}$ and divacancy complexes increase to at% concentrations which we believe would have been observed if actually present in real-world samples.

Thus, at least for the case of heavily Sn-doped EFG grown crystals, the real-world temperature dependencies of chemical potentials and bandstructure are absolute requirements for quantitative defect modelling. The role of $\Delta S_{vib}$ is also clearly very important, but we hesitate to draw absolute conclusions because of the oversimplicity of our treatment herein; we eagerly await developments in efficient but high-fidelity computation of defect vibrational spectra. Versions of Fig. 5 (a) and (c) for a range of f from 0 to 1 are presented in the Supplemental Materials.

*Modeling Unintentionally-Doped (UID), Fe-Doped, and Mg-Doped Crystals*

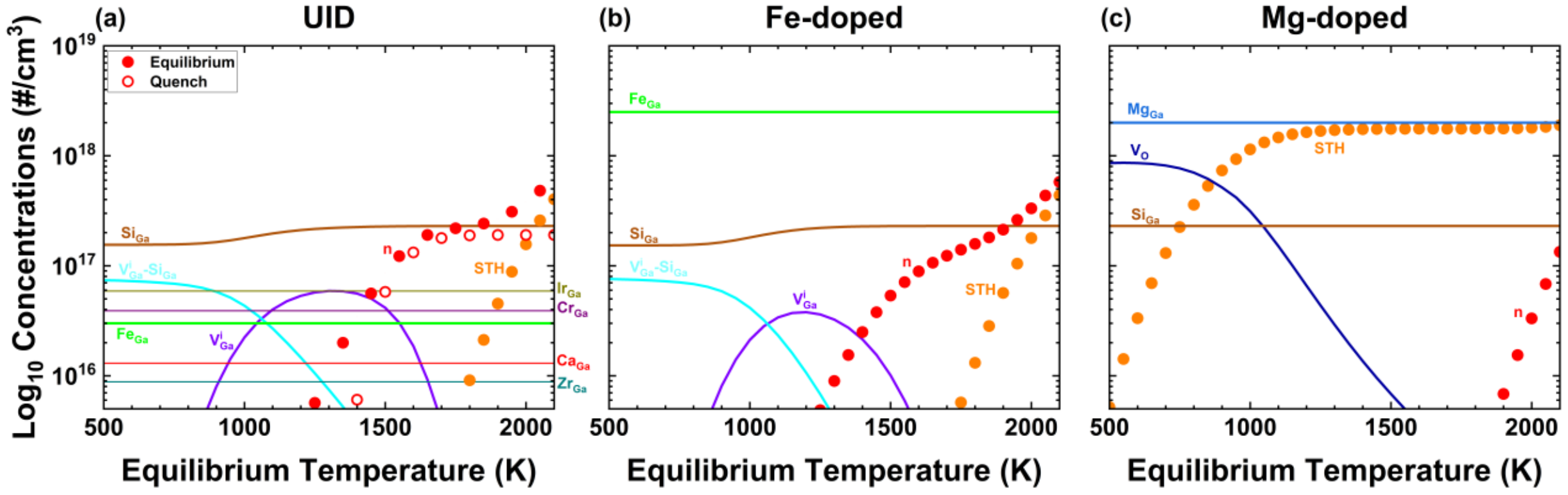

**Figure 6.** Calculated defect concentrations for EFG grown ($p_{O2}$=0.02 atm) (a) UID, (b) Fe-doped $Ga_2O_3$ and (c) Mg-doped $Ga_2O_3$. Background impurities have concentrations as in Fig. 1 (a), while [Fe] and [Mg] were set to 3 x$10^{16}$ /cm$^3$ and 5 x$10^{15}$ /cm$^3$. The predicted carrier densities are in good agreement; n=[Si] for UID while immeasurable carrier densities should be present for Fe and Mg doped which is consistent with them being insulating.

Figure 6 investigates defects predicted for cases of unintentionally-doped (UID), Fe-doped and Mg-doped β-$Ga_2O_3$ grown by EFG, with non-dopant-related model parameters fixed at the values determined in Fig. 1(a) unless otherwise noted. The results agree with real-world observations that UID are conducting while Fe and Mg-doped β-$Ga_2O_3$ are insulating. In Fig. 6 (a), the UID case is assumed to have dominant impurities [Si]=2.3x$10^{17}$ /$cm^3$, [Mg]=5x$10^{15}$ /$cm^3$ and [Fe] = 3x$10^{16}$ /$cm^3$ [33], and a wide range of temperatures yield equilibrium n = [Si] down to 1700 K. Note that equilibrium n and STH concentrations exceed the low, unintentional [Si] at the highest temperatures; since both electrons and STHs can equilibrate from any processing temperature this is of no consequence at room temperature. In (b), [Fe]=2.5x$10^{18}$ /$cm^3$ and [Si]=2.3x$10^{17}$ /$cm^3$ for all temperatures while in (c) [Mg]=2x$10^{18}$ /$cm^3$ and [Si]=2.3x$10^{17}$ /$cm^3$ [1,51]. Due to the lack of detailed data on other unintentional impurities in such crystals, (b) and (c) can be considered as hypothetical, assuming that the inclusion of Ir and other impurities remain constant. For Fe-doped (b) and Mg-doped (c), electron concentrations fall to immeasurable values by 300 K for both cases whether cooled slowly or quenched, thus both would be insulating for any cooling rates. The main difference compared to the nearly-degenerately Sn-doped cases above is the greater role played by STHs because their numbers either exceed the doping in the case of UID, or because of the acceptor dopants in the Fe and Mg cases. Thus, our modelling reproduces the major known aspects of defect quantities in Sn-, Fe-, Mg- and UID EFG-grown crystals. In the Supplementary Materials, we additionally model Bridgeman growth in Pt/Rh crucibles at $pO_2$=1 atm. Follow-on modelling of epitaxially-grown crystals by various methods would help to provide additional constraints on our quantitative defect modelling.

## Conclusions

We quantitatively modeled point defect concentrations in β-$Ga_2O_3$ by combining formation energies from DFT using hybrid functionals with advanced thermodynamic modelling using a new program we dub KROGER. We find that, to achieve agreement with real-world experience that Sn-doped wafers are conductive with n=[Sn], many oft-neglected temperature-dependent effects must be included and that constant concentration rather than constant-chemical potential thermodynamic conditions are more appropriate for describing [Sn]. Extensive thermochemical data is available for many semiconductors; using this data to compute temperature- and pressure-dependent chemical potentials allows high-fidelity modelling of specific growth and processing. Including $E_g(T)$ is also critical because it suppresses $V_{Ga}$ that would otherwise strongly compensate the n-type doping. We included a simple, minimal estimate of defect $\Delta S_{vib}$ based on counting quantum oscillator modes per atom which is especially important for divacancies and other large complexes. Especially for oxides like β-$Ga_2O_3$ for which crystal growth occurs at very high temperatures, $T\Delta S_{vib}$ amounts to 1-2 eV thus while computationally-demanding calculations are required, accurately computing this factor will be critical to advancing defect modelling. We have demonstrated a framework and workflow for high-fidelity modelling of defect concentrations in semiconductors, which we anticipate will be extremely useful in moving defect computations from qualitative insights to quantitative process-dependent predictions.

## Calculation Methods

*Formation Free Energy of Defects*

The formation energy of isolated, dilute defects in crystals can be computed under either isobaric (Gibbs energy) or isochoric (Helmholtz energy) constraints. The total free energy of the $q^{th}$ charge state of the $j^{th}$ defect or complex is given as:

$$E_{j,q} = E_{j,q}^{o} + q\Delta\varepsilon_F - \sum_i \Delta\mu_i m_{i,j} - T\Delta S_{j,q}^{vib} \quad (2)$$

in which $E_{j,q}^{o}$ is the finite-size corrected formation energy in the dilute limit for a Fermi level ($\varepsilon_F$) evaluated at the position of the valence band maximum ($\varepsilon_v$) [52] herein calculated using DFT with Heyd-Scuzeria-Ernzerhof (HSE)[53,54] hybrid functionals, $\Delta\varepsilon_F$ represents the Fermi energy (chemical potential for electrons) relative to the electrochemical energy reference ($\varepsilon_F - \varepsilon_v$) used for $E_{j,q}^{o}$, $\Delta\mu_i$ is the chemical potential of element *i* referenced to its standard state, $m_{i,j}$ is the number of *i* atoms added (+) or removed (-) to form the $j^{th}$ defect or complex, and $\Delta S_{j,q}^{vib}$ is the associated change in vibrational entropy of the crystal. $\Delta S_{j,q}^{vib}$ may be sensitive to isochoric versus isobaric conditions for anharmonic bonding and remains computationally demanding to compute accurately for isolated defects at useful levels of theory (e.g. using HSE and very large supercells).

*Defect concentrations*

In the dilute limit, the number of each charge state $q$ of defect $j$ ($n_{j,q}$) is given by:

$$n_{j,q} = \theta_{j,q} N_{j,q}\, exp\left(-\frac{E_{j,q}}{k_B T}\right) \quad (3)$$

We use the number density of appropriately-sized unit cells $N_{j,q}$ as the basis for each defect or complex since some large complexes may require multiple primitive cells (or equivalently, lattice sites or formula units). For example, we consider complexes containing up to $3Sn_{Ga}$ donors bound to a $V_{Ga}$, which would require between 2 and 4 formula units depending on whether each of these 4 elementary defects entities occupy $Ga_1$ or $Ga_2$ sites. In other words, the numerical prefactor commonly taken as $N_{sites}$ should vary for large complexes. Additional degeneracy factors for configurational, electronic, and spin degrees of freedom are combined into $\theta_{j,q}$ [55]. All calculations in this paper included a set of 873 charge states comprising 259 defects and complexes and 19 elements in β-$Ga_2O_3$. Most (but not all) have been previously published in Varley et al., [8–10] and Frodason et al., [4–6]; the formation energies are provided for all of these defects in the Supplementary Materials. The formation energy of β-$Ga_2O_3$ was calculated for the same supercells as -10.22 eV/formula unit (FU), which is close to but lower than the experimentally measured value at 300 K.

We also implemented self trapped holes (STH) as new categories of electronic defects in addition to the usual free electrons and holes whose numbers are calculated from effective conduction band and valence band densities of states, respectively. The self-trapping energies for STH on $O_I$ and $O_{II}$ sites are taken as 0.53 & 0.52 eV [30,31]. An important detail is that, in thermal equilibrium, both band holes and self-trapped holes are equally-accessible microstates for electron-hole pair excitation of the perfect crystal. Since every $O_I$ and $O_{II}$ atom is capable of localizing a STH and in the fully-localized limit they will form a dispersionless band, the prefactor for each STH type is the primitive unit cell density $1.91 \times 10^{22}$ /cm$^3$ rather than the effective valence band density of states = $1.71 \times 10^{20}$ /cm$^3$ we adopt herein for band holes[56]. While optical excitations measure the "bandgap" of β-$Ga_2O_3$ to be 4.8-5 eV (strongly modified by Urbach tails [3]), from the standpoint of calculating the intrinsic carrier density the "bandgap" is arguably 0.53 eV lower because thermal excitations can create STH's while photons cannot. Because the intrinsic carrier density including STH ($n_{i,STH}$) approaches ~$10^{17}$ /cm$^3$ near $T_{melt}$, this slightly affects defect equilibrium at high temperatures but does not change any major findings herein although it may for cases of lower impurity doping concentrations.

*Charge Neutrality and Quenched Concentrations*

Within the usual grand-canonical formalism, the simplest defect equilibrium problem is full equilibrium at a given temperature with chemical potentials specified for all elements. Each charge state included in modeling adds one unknown concentration but also one equation of the form Eq. 3. However, $E_{j,q}$ for nonzero charge states depends on the additional unknown $\varepsilon_F$, thus the final equation allowing implicit solution of the system of equations comes from charge balance, e.g. solving the charge-neutrality condition:

$$0 = STH_I + STH_{II} + p - n + N_D^+ - N_A^- + \sum_{j,q} q\, n_{j,q} \qquad (4)$$

in which provision is made for anonymous ionized donors ($N_D$) and acceptors ($N_A$), and the final term is the total charge in all the charge states included in the model. The conduction electron (n), valence band hole (p), and STH densities are computed using Fermi-Dirac statistics.

Equation 3 can be overridden with fixed values if the concentrations of some charge states should be held constant. If the total concentration $n_{j,tot}$ of a defect j composed of multiple charge states (given by $n_{j,tot} = \sum_q n_{j,q}$) should be held constant, this is accomplished by overriding Eq. 3 with values from the Gibbs distribution:

$$n_{j,q} = n_{j,tot}\, exp\left(-\frac{E_q}{k_B T}\right) / \sum_q exp\left(-\frac{E_q}{k_B T}\right) \qquad (5)$$

This allows modelling temperature dependent defect concentrations in cases where certain defects stop equilibrating below some temperature but the charge in defect j still changes with temperature.

Full quenching a system of defects from a processing temperature to a final temperature (usually 300 K) is accomplished by using Eq. 5 for all defects in the model with the $n_{j,tot}$ determined at the high temperature. The total concentration of impurity i contained in all defects in the model is of course simply the number of i atoms per defect j times the total defect j concentration, summed over all defects in the model:

$$n_i = \sum_j n_{j,tot} \cdot m_{i,j} \qquad (6)$$

For the host material elements, this sum must be added to the numbers of atoms in non-defective unit cells.

*Specification of Chemical Potentials or Concentrations for Elements*

In a grand canonical formulation of point defect formation, it is assumed that atoms are exchanged between the crystal and reservoirs setting chemical potentials for each element until equilibrium is reached. It is common (but not universal) in DFT-based work on defects to compute temperature-independent formation energies of all phases also using DFT and then evaluate temperature dependent defect concentrations in rich/poor chemical potential limits for each element. Using temperature-independent chemical potentials assumes that the specific heats of all relevant compounds are negligible. For gaseous reactants, this is corrected for the ideal gas translational kinetic energy contribution $\Delta\mu(T,P) = \mu_{ref} + k_b T \ln\left(\frac{p}{p_{oref}}\right)$, but for multiatomic molecules and high temperatures, the contributions of rotovibrational and electronic excitations also contribute. These can be challenging to compute accurately, so we adopted a pragmatic approach recognizing that for many materials of interest, experimental temperature- and pressure-dependent thermochemical data such as standard Gibbs energies ($G^o(T,P)$) are available that have been validated to reproduce the materials' phase diagrams. Zinkevich and Adlinger assessed the thermodynamics of the Ga-O binary system (referenced to 0 K) and we adopt their temperature-dependent parameterizations[35]. Values for competing phases involving impurities were taken from the Fact Sage database[57]. The fact that oxygen is a gas for all growth conditions means that temperature and partial pressure significantly change its chemical potential

$$\mu_O = \frac{\mu_{O_2}}{2} = \frac{G°_{O_2}(T,P°)}{2} + \frac{k_B T}{2}\left[ln\left(\frac{P_{tot}}{P°}\right) + ln\left(\frac{P_{O_2}}{P_{tot}}\right)\right] \quad (6a)$$

$$\Delta\mu_O = \mu_O - \mu°_O = \frac{G°_{O_2}(T,P°) - G°_{O_2}(T°,P°)}{2} + \frac{k_B T}{2} ln\left(\frac{P_{O_2}}{P^o}\right) \quad (6b)$$

in which $G^{\circ}_{O_2}(T, P^{\circ})$ is the standard Gibbs energy as function of temperature, $P^o$ is the reference pressure (e.g. 1 atm), $p_{O2}$ is the oxygen partial pressure, and in Eq. 6a we have avoided cancelling out $P_{tot}$ to clarify the roles of total and partial pressures (which can become confusing especially when equilibrium vapors such as $Ga_2O$ + $O_2$ resulting from $Ga_2O_3$ decomposition are considered [35,36,58]). The temperature dependencies from rotovibrational and electronic degrees of freedom are contained in $G°_{O_2}(T, P°)$. If the reference temperature for the thermochemical data is not 0 K then the defect formation energy from the elements should be corrected by $\Delta G = \int_0^{Tref} \Delta C_p(T)\left(1 - \frac{1}{T}\right) dT$. For ternary and more complex materials and for novel materials, this data may not exist thus computation of the formation free energy is required [59].

Each growth or annealing process we considered occurs under well-defined temperature, total pressure, and oxygen partial pressure ($p_{O2}$), allowing us to determine the excess chemical potential for oxygen $\Delta\mu_O$ in Eq. 6b in equilibrium with β-$Ga_2O_3$. Calculations may be made for values of chemical potentials anywhere within the boundaries of the β-$Ga_2O_3$ single phase field. At high temperatures, β-$Ga_2O_3$ may be in equilibrium with $Ga_{(vap)}$, $GaO_{(vap)}$ or $Ga_2O_{(vap)}$ on the Ga-rich side and with $O_{2(g)}$ on the O-rich side, depending on the experiment. For all growth processes herein, $pO_2$ is controlled directly so $\Delta\mu_{Ga}$ is determined from the formation reaction

$$2Ga + \frac{3}{2}O_2 \rightleftharpoons Ga_2O_3 \quad (7a)$$

leading to

$$2\Delta\mu_{Ga} + 3\Delta\mu_O = G^{\circ}_{Ga2O3} \quad (7b)$$

Only for $pO_2$ lower than all cases herein does decomposition into $Ga_2O$ and $O_2$

$$Ga_2O + O_2 \rightleftharpoons Ga_2O_3 \quad (7c)$$

become a limiting reaction. To model specific growth processes, we carry out calculations with either the concentration or chemical potential specified for each element, representing whether or not it equilibrates with the growth environment. This is determined by kinetics like surface exchange or diffusion and thus will be dependent on dimensions and cooling rates. In polycrystals or thin films heated or cooled slowly, interstitial diffusing elements are better described via chemical potentials, while substitutional diffusers in rapidly cooled bulk crystals may be better described by fixed concentrations [60,61]. Maximal values of chemical potentials are set by equilibria with 2nd phases, but lower chemical potentials or concentration may be fixed by the particular growth process; for example when a dopant like Sn is supplied at concentrations lower than the solubility limit at $T_{melt}$ for liquid phase growth. Thus, process modelling of defects requires augmenting the typical grand canonical approach with the ability to simultaneously solve for the Fermi level and for chemical potentials of elements of fixed concentration.

*Vibrational Entropy*

In lieu of directly computing the vibrational energy changes associated with each chargestate, we adopt a simple quantum oscillator model[62] to calculate the vibrational entropy ($\Delta S^{vib}_{j,q}$) using one representative phonon frequency $\omega_o$ for the solid rather than the detailed phonon band structure[55]. Each atom in the perfect lattice adds 3 degenerate modes at $\omega_o$; therefore forming a vacancy subtracts 3, an interstitial adds 3, a substitutional defect does not change the number of modes, and complexes add or subtract modes according to the net change in number of atoms. Because of our lack of information regarding bond stiffness changes, herein we only count changes in numbers of "average" modes rather than taking into account mode frequency changes. Such information can readily be incorporated, as can integration over phonon bandstructures for perfect and defective supercells.

Within our simple treatment, the vibrational entropy and isochoric specific heat per atom are:

$$S_{vib,atom} = -3k_B\left[\frac{\hbar\omega_o}{2k_BT}coth\left(\frac{\hbar\omega_o}{2k_BT}\right) + ln\left(\frac{1}{2}csch\left(\frac{\hbar\omega_o}{2k_BT}\right)\right)\right] \quad (8)$$

$$C_v = \frac{\hbar^2\omega_o^2}{4k_BT^2}csch^2\left(\frac{\hbar\omega_o}{2k_BT}\right) \quad (9)$$

in which $\omega_o$ is the 'average' phonon frequency, which we determined by setting the Debye temperature of the associated phonon-only specific heat to 872 K [63], $k_B$ is Boltzmann's constant, $T$ is absolute temperature and $\hbar$ is the reduced Planck's constant.

Mass and bonding changes will indubitably result in different magnitudes and even different signs of $\Delta S_{vib}$ [64,65] for different charge states. Improved calculations of $\Delta S_{vib}$ using hybrid functionals or higher levels of theory for dominant defects and complexes is one of the final frontiers for refinement of the modelling herein. We note that crystal growth experiments are nearly always done at controlled by pressure rather than volume; thus computations should reflct consistent assumptions. The difference between the two is related to thermal expansion and constant pressure conditions would in general tend to lower phonon frequencies since interatomic energy vs. distance curves soften at larger distances. Thus our treatment can also be improved by computing $\Delta S_{vib}$ for different lattice constants taking into account (anisotropic) thermal expansion.

*Temperature Dependence of Bandgap*

Both bandgap and effective densities of states are strong functions of temperature; many semiconductors exhibit on the order of about -0.1 eV per 300 K decrease in $E_g(T)$, which can be parameterized using e.g. the Varshini or Einstein-Debye equations[37,66] (we used the values from Lee et. al). Because the fractions of bandgap change from the conduction and valence bands are not definitively known for $\beta$-$Ga_2O_3$, we introduce the parameter f representing the fraction of $\Delta E_g(T)$, arising from the conduction band [1,37,67]. The effective density of states for isotropic, parabolic bands has $T^{3/2}$ dependence. For generality we utilize Fermi-Dirac statistics for band carriers and also can include degenerate statistics for site occupation by defects[68] if required. Based on computed effective masess for band electrons and holes, we adopted values of $N_C=3.33\times10^{18}$ /cm$^3$ and $N_V=1.71\text{x}10^{20}$ /cm$^3$ [56] at 300 K and the bandgap at 0 K as $E_g=5.1$ eV[69]

*Solution Methods*

For full equilibrium calculations in which all elements are specified by chemical potentials, only Eq. 4 must be solved; KROGER's default method is Matlab's fzero function. An initial $\varepsilon_F$ guess is generated by computing the net charge on a grid spanning from $E_v$ to $E_c$, plus 5 $k_BT$ on either side and taking the point where the sign changes. The direct search is done because commonly the net charge barely changes with $\varepsilon_F$ over much of the bandgap so traditional downhill optimization methods may stall out unless good initial guesses close to the solution are given. Calculating defect equilibrium with some element concentrations specified is more challenging as it requires simultaneously solving for the $\varepsilon_F$ and corresponding chemical potentials (e.g. $\Delta\mu_{Sn}$ herein) that satisfy charge balance and yield the specified total element concentration across all defects. In KROGER this is achieved by minimizing a composite objective function composed of the absolute value of net charge (right side of Eq. 4) plus the absolute value(s) of the deviation(s) of concentration(s) of the fixed element(s) from their target number(s) (for each element, Eq. 6 minus the target value). Because of the exponential dependence of concentrations on $\varepsilon_F$ and chemical potentials, the solution is typically located in extremely narrow hypervalleys of width $<k_BT$ surrounded by large hyperareas having ~0 gradient; thus self-consistent solution requires rather exhaustive searches of the parameter space. KROGER retains a brute-force grid search option but we have found that particle swarm optimization followed by a traditional simplex optimization balances accurate convergence with speed. Calculations versus temperature are done efficienctly by exploiting continuity by proceeding from high to low temperature using the prior solution as the guess for $\varepsilon_F$ and $\Delta\mu$. Solutions are accepted when the composite objective function is of order $10^3$-$10^6$ /cm$^3$, which is at least 6-8 orders of magnitude less than practically-measurable concentrations.

*Uncertainty Estimates*

Despite thorough optimization with the model described above, some uncertainties in parameter values remain. Based on sensitivity analysis and the magnitude of free energy terms, we estimate the uncertainties in $T\Delta S_{vib}$ at $T_m$ for each charge state to be $\pm$1 eV because of our crude approximate model, the uncertainty in the thermal bandgap energy $E_g$ to be $\pm$0.3 eV (including the issues of STHs mentioned) and the uncertainties related to the HSE-calculated charge state formation energies to be $\pm$0.5 eV [49]. We have carefully assessed that the impact, even when these are compounded, on the major findings herein is minimal. For example electron density n at 1300-2100 K changes by only a small factor 0.5-2x for Sn-doped samples because of the constraint of charge balance. Addressing these uncertainties represents frontiers of defect computation and will lead to even more accurate predictions.

## Conflicts of interest

There are no conflicts to declare.

## Code and Data Availability

The defect properties including HSE-computed formation energies are available in the Supplementary Materials. The KROGER code is available at https://github.com/mikescarpulla/KROGER

## Acknowledgements

This work was supported by funding from the Air Force Office of Scientific Research under MURI Award No. FA9550-21-0078 (Program Manager. Dr. Ali Sayir). The work of J.B.V. was performed under the auspices of the US DOE by Lawrence Livermore National Laboratory under contract DE-AC52-07NA27344 and partially supported by LLNL LDRD Funding under Project 22-SI-003 and by the Critical Materials Institute, an Energy Innovation Hub funded by the U.S. DOE, Office of Energy Efficiency and Renewable Energy, Advanced Materials and Manufacturing Technologies Office. Financial support is acknowledged from the Research

Council of Norway through the GO-POW project (Grant No. 314017). Computations were performed on resources provided by UNINETT Sigma2 -- the National Infrastructure for High Performance Computing and Data Storage in Norway.